# Quantitative three-dimensional absorption imaging in standard brightfield microscopes

Yoonjae Chung[1,2,7+], Sehyeon Lee[2,3+], Herve Hugonnet[2,3], Chulmin Oh[2,3], Wei Sun Park[2,3], Yeon Wook Kim[4], Seung-Mo Hong[5], YongKeun Park[2,3,6*]

[1]Department of Electrical Engineering, KAIST, Daejeon, 34141, Republic of Korea

[2]KAIST Institute for Health Science and Technology, KAIST, Daejeon, 34141, Republic of Korea

[3]Department of Physics, KAIST, Daejeon, 34141, Republic of Korea

[4]Biomedical Research Center, Asan Institute for Life Sciences, Asan Medical Center, Seoul, Republic of Korea

[5]Department of Pathology, Asan Medical Center, University of Ulsan College of Medicine, Seoul 05505, Republic of Korea

[6]Tomocube Inc., Daejeon 34109, Republic of Korea

[7]Current affiliation: Department of Electrical Engineering, Stanford University, Stanford, CA 94305, United States

[+]Equally contributed

[*]E-mail: yk.park@kaist.ac.kr

**Abstract**

Optical absorption is a primary, label-defining contrast across biology, pathology, and materials science, yet three-dimensional quantitative absorption imaging has remained largely inaccessible to the brightfield microscopes used in everyday practice. We introduce quantitative absorption tomography (QAT), which recovers volumetric distributions of the extinction coefficient by treating brightfield image formation as a linear inverse problem in logarithmic intensity space and inverting a three-dimensional absorption optical transfer function. Under weak-scattering conditions, QAT yields spectrally resolved, three-dimensional absorption maps from through-focus image stacks acquired on standard brightfield platforms, without interferometry, coherent illumination, or sample rotation. We use QAT to track melanin dynamics in living melanoma cells without exogenous labels, image pigment organization in intact Petunia hybrida petals in vivo, and reconstruct chromogenic contrast across large H&E-stained human tissue volumes. By establishing absorption as a directly measurable volumetric quantity within standard brightfield workflows, QAT positions chromogenic contrast as a quantitative axis alongside fluorescence- and refractive-index-based imaging.

**Introduction**

Optical absorption underlies the most widely used imaging contrasts in biology and pathology, from chromogenic stains and immunohistochemistry to endogenous pigments such as melanin and hemoglobin. Yet three-dimensional quantitative absorption imaging is not routinely available on the brightfield microscopes that produce these contrasts. Here we close this gap by treating brightfield image formation as a linear inverse problem in logarithmic intensity space, recovering the three-dimensional extinction coefficient from a single focal stack of a standard brightfield microscope.

Brightfield microscopy remains one of the most widely used imaging modalities in biology and medicine owing to its simplicity, robustness, and compatibility with chromogenic stains[1]. Histopathology based on hematoxylin and eosin (H&E) staining and brightfield microscopy constitutes the gold standard for disease diagnosis and guides clinical decision-making across a broad range of conditions. Despite its central role, conventional brightfield microscopy is limited to two-dimensional (2D) observation of thin tissue sections, typically 4–5 µm thick. Standard brightfield imaging also provides only qualitative, 2D absorption contrast, inferred from transmitted intensity, rather than quantitative measurements of the absorption coefficient. As a result, critical three-dimensional (3D) information remains largely inaccessible in a systematic and quantitative manner[2–4].

Recovering volumetric information from biological specimens has long been a central objective of optical microscopy and its application to biology. Recent advances in optical tomography have further highlighted the potential of slide-wide, high-throughput volumetric imaging for clinical diagnostics[5]. Traditionally, three-dimensional reconstruction has relied on serial sectioning followed by digital registration, but this workflow is labor-intensive, destructive, and susceptible to section loss, mechanical distortion, and alignment errors[6]. Optical alternatives, including fluorescence-based volumetric imaging and tissue clearing combined with light-sheet microscopy[7], enable high-resolution 3D visualization but require specialized labeling, sample preparation, and instrumentation that are not routinely available in clinical pathology workflows[8]. Fluorescence contrast also differs from chromogenic absorption contrast, complicating direct comparison with established histological standards.

A number of optical strategies have sought to extend brightfield microscopy into three dimensions. Optical projection tomography and related methods reconstruct volumetric attenuation by acquiring angular projections[9,10], but typically require sample rotation or optical clearing and suffer from resolution trade-offs at high numerical apertures. Other approaches improve axial sectioning through computational deconvolution of brightfield focal stacks[11,12], yet these methods are often qualitative and limited to serial ultrathin sections. They do not yield quantitative absorption coefficients or spectral separation[12–14]. As a result, a general, quantitative framework for 3D absorption imaging compatible with standard brightfield microscopes has been lacking. Recent transport-of-intensity diffraction tomography (TIDT) methods reconstruct refractive index volumes from brightfield intensity stacks acquired under tailored illumination[15–17]. Structured-illumination approaches can recover joint phase and absorption maps in two dimensions[18,19]. However, absorption itself has not been treated as a primary three-dimensional reconstruction target.

Several label-free optical techniques have been proposed to infer 3D tissue structure without serial sectioning. Holotomography, a 3D quantitative phase imaging technique, reconstructs the refractive index (RI) distribution of thick specimens, exploiting RI as an intrinsic imaging contrast[20,21]. While powerful for transparent objects, these approaches rely on indirect inference of chromogenic contrast and require careful validation against ground truth. Although joint reconstruction of phase and absorption has been demonstrated within quantitative phase imaging (QPI) frameworks[22,23], direct, quantitative, and color-resolved 3D absorption imaging of stained specimens has remained comparatively unexplored.

Here, we introduce QAT, a computational framework that reconstructs 3D absorption (extinction) coefficient distributions from conventional brightfield focal stacks. The central insight of QAT is that, under weak-scattering conditions, absorption can be modeled as a linear imaging process when expressed in logarithmic intensity space. This formulation enables computation of an absorption optical transfer function (OTF) and its inversion via deconvolution, yielding quantitative volumetric absorption maps without interferometry, coherent illumination, or sample rotation. In its simplest implementation, QAT requires only standard Köhler illumination and axial scanning, making it compatible with existing brightfield microscope platforms.

We demonstrate applicability across a wide range of specimens and length scales. Using spectrally selective absorption phantoms, we validate accurate 3D localization and wavelength-specific absorption reconstruction. We show that QAT provides absorption-specific contrast that complements RI tomography in pigment-containing living melanoma cells and enables long-term, label-free monitoring. We further extend QAT to in vivo imaging of *Petunia hybrida* petal cells, resolving endogenous pigment distributions. Finally, we apply QAT to large-area, H&E-stained human rectal tissue sections. By enabling quantitative, 3D absorption imaging using brightfield microscopy, QAT bridges a critical gap between conventional brightfield imaging and volumetric optical imaging. This approach offers a practical pathway toward 3D biology, histopathology, and spatial analysis of stained and pigmented specimens using familiar imaging workflows.

## Results

### Image formation and inverse model

Microscopic image formation can be described by the interaction of a complex RI distribution $\bar{n}$ of the sample $\bar{n}(x,y,z) = n(x,y,z) + i\kappa(x,y,z)$ with the illumination. The real RI values $n$ describe light refraction (locally changing illumination's phase) while the imaginary RI $\kappa$ —the extinction coefficient— relates to light absorption (locally changing illumination's strength). Under the assumption of a weakly scattering object, a linear relation between the measured image intensity attenuation $S = \log[I(\mathbf{r})/I_0(\mathbf{r})]$ and the object complex RI values can be derived[24]:

$$S = H_A * V_{imag} + H_P * V_{real},$$

where $V_{imag}$ and $V_{real}$ are the imaginary and real parts of the scattering potential $V = (2\pi/\lambda)^2[\bar{n}^2(\mathbf{r}) - n_0^2]$, $\lambda$ is the wavelength and $n_0$ is the surrounding background RI. Under the weak scattering condition ($\bar{n} - n_0 \ll 1$), the extinction coefficient and RI are proportional to the imaginary and real part of the scattering potential $\kappa \cong \frac{V_{imag}}{8\pi^2/\lambda^2 n_0}$ and $n \cong \frac{V_{real}}{8\pi^2/\lambda^2 n_0}$, making reconstruction of either quantity equivalent within the linearized model. Throughout this work, $\kappa$ denotes the extinction coefficient (imaginary part of the RI) rather than the absorption coefficient $\mu_a$, with the relation $\mu_a = 2k_0\kappa$, where $k_0 = 2\pi/\lambda$ is the vacuum wavenumber.

The validity of this linearized absorption model requires that absorption- and phase-induced intensity modulations remain sufficiently weak, analogous to the linearized forward models commonly employed in transport-of-intensity and intensity-based diffraction tomography. In such intensity-based tomographic formulations, quantitative validity bounds on optical thickness and accumulated phase delay are required to ensure the accuracy of the linear approximation[25]. Specifically, the optical thickness associated with absorption must satisfy $\kappa \cdot L \ll 1$, and the accumulated phase delay should remain below approximately one radian, $|\Delta\varphi| =$ $(2\pi/\lambda)\Delta n \cdot L \leq 1$, where $\Delta n = \bar{n} - n_0$ and $L$ denotes the refractive index contrast and the sample thickness, respectively. Under these conditions, multiple scattering and higher-order coupling between phase and absorption can be neglected, justifying the logarithmic linearization used in QAT (see Supplementary Note 1).

$\mathrm{H_A}$ and $\mathrm{H_P}$ are the absorption and phase transfer function of the microscope which can be computed from the pupil plane intensity distribution $\rho$ as follows[26],

$$H_P = -128\pi^4 Im[G^* \mathcal{F}^{-1}\{\tilde{G} k_z \rho\}],$$
$$H_A = -128\pi^4 Re[G^* \mathcal{F}^{-1}\{\tilde{G} k_z \rho\}],$$

where $k_z$ is the axial wavevector component, $\mathcal{F}$ the Fourier transform and the tilde $\tilde{}$ denotes Fourier transformed quantities, and $G(x,y,z) = \frac{i}{8\pi^2} \iint_{\sqrt{k_x^2+k_y^2}<NA} \frac{e^{ik_x x + ik_y y + ik_z z}}{k_z} dk_x dk_y$ denotes the green's function truncated to the numerical aperture of optical system. $\boldsymbol{k} = (k_x, k_y, k_z)$ denotes the wavevector. The extinction and refractive index images of the sample are obtained through deconvolution of the microscope intensity stack by the transfer function.

To separate the contribution of $V_{imag}$ and $V_{real}$ at least two measurements $S_i$ with different illuminations $\rho_i$ are needed and image formation is expressed with a matrix equation[27].

$$\tilde{\boldsymbol{S}} = \begin{pmatrix} \tilde{S_1} \\ \tilde{S_2} \\ \cdots \end{pmatrix} = \begin{pmatrix} \tilde{H}_{A1} & \tilde{H}_{P1} \\ \tilde{H}_{A2} & \tilde{H}_{P2} \\ \cdots & \cdots \end{pmatrix} \begin{pmatrix} \tilde{V}_{imag} \\ \tilde{V}_{real} \end{pmatrix} = \tilde{\boldsymbol{H}} \tilde{\boldsymbol{V}},$$

which is solved by Wiener deconvolution with regularization $\alpha$:

$$\tilde{\boldsymbol{V}} = (\tilde{\boldsymbol{H}} \tilde{\boldsymbol{H}}^\dagger + \alpha)^{-1} \tilde{\boldsymbol{H}}^\dagger \tilde{\boldsymbol{S}}.$$

When $V_{real}$ is not of interest, one can instead determine $V_{imag}$ with a single illumination if $H_P = 0$. This condition is met for an illumination $\rho = 1/k_z$ since $G^* \mathcal{F}^{-1}\{\tilde{G} k_z \rho\} = |G|^2$ resulting in $H_P = 0$. However, in realistic biological specimens, residual phase–absorption coupling may persist due to RI heterogeneity and partial coherence effects. Such phase–absorption coupling has been shown to introduce cross-talk and reconstruction bias in intensity-based diffraction tomography[17,23,28], motivating the use of multi-illumination schemes for robust decoupling of absorptive and refractive contributions.

The above procedure separates $V_{imag}$ and $V_{real}$, from which $\kappa$ and $n$ are computed. Unlike conventional brightfield deconvolution approaches that aim to suppress out-of-focus blur without assigning physical meaning to the reconstructed intensity, QAT explicitly links the reconstructed volume to the absorption coefficient through a quantitative inverse model derived from 3D transfer function theory. In practice, QAT suppresses phase-induced intensity variations by exploiting illumination symmetry and transfer-function properties, enabling absorption-dominant volumetric contrast under weak-scattering conditions.

**Validation with spectrally selective phantoms**

Spectrally selective absorption phantoms enable wavelength-resolved 3D validation of QAT against a well-defined ground truth. QAT relies on focus stack acquisition and reconstructs quantitative absorption coefficients by explicitly computing the point spread function from 3D OTF theory, accounting for both refraction and absorption effects[25,29].

In our implementation (Fig. 1a), programmable angular illumination was achieved using a digital micromirror device (DMD) positioned at the condenser pupil plane. The DMD projected sequential illumination patterns under red (624 nm), green (520 nm), and blue (455 nm) LED illumination, while transmitted light was collected by a high–numerical-aperture objective. Axial scanning of the sample generated wavelength-resolved brightfield focal stacks encoding both in-focus and out-of-focus absorption contributions.

To quantitatively evaluate the performance of QAT, we imaged a mixture of cyan, magenta, and yellow (CMY) laser printer toner particles, which selectively absorb red, green, and blue light. This sample serves as a spectrally orthogonal absorption phantom with wavelength specificity. Following deconvolution of the logarithmic intensity stacks using wavelength-specific absorption transfer functions, QAT reconstructs three-dimensional absorption volumes with strong suppression of out-of-focus background and accurate spatial localization.

In Figure 1b, orthogonal slices precisely resolve the individual particles along the axial direction, with minimal elongation compared to the raw brightfield data. QAT preserves spectral selectivity: cyan, magenta, and yellow particles are correctly assigned to their respective absorption channels, with negligible axial color bleed-through. These results demonstrate that QAT not only enhances axial sectioning but also faithfully recovers wavelength-dependent absorption contrast in three dimensions.

Although multi-illumination QAT enables simultaneous reconstruction of absorption and refractive index (Fig. 1c), absorption-only QAT can be performed using only one $\rho = 1/k_z$ illumination pattern. In this mode, all measurements were acquired using eight sequential DMD-generated pupil patterns that were summed to form the single refraction insensitive pattern. This strategy maintains consistency with acquisition protocols used for joint phase–absorption reconstructions in live-cell experiments while demonstrating the single-pattern reconstruction capability.

In this work, we employ both modes depending on the application. Multiple structured illuminations are used to jointly reconstruct refractive index and absorption in live-cell experiments, while absorption-only QAT is adopted for large-area imaging of stained tissue to maximize robustness and throughput. In all cases, reconstructed extinction coefficient volumes are further refined using total variation regularization with channel-specific weights ( $\lambda_{\mathrm{TV},R} = 0.001$ , $\lambda_{\mathrm{TV},G} = 0.004$ , $\lambda_{\mathrm{TV},B} = 0.001$ ) and non-negativity constraints[30], improving noise robustness while preserving sharp axial sectioning.

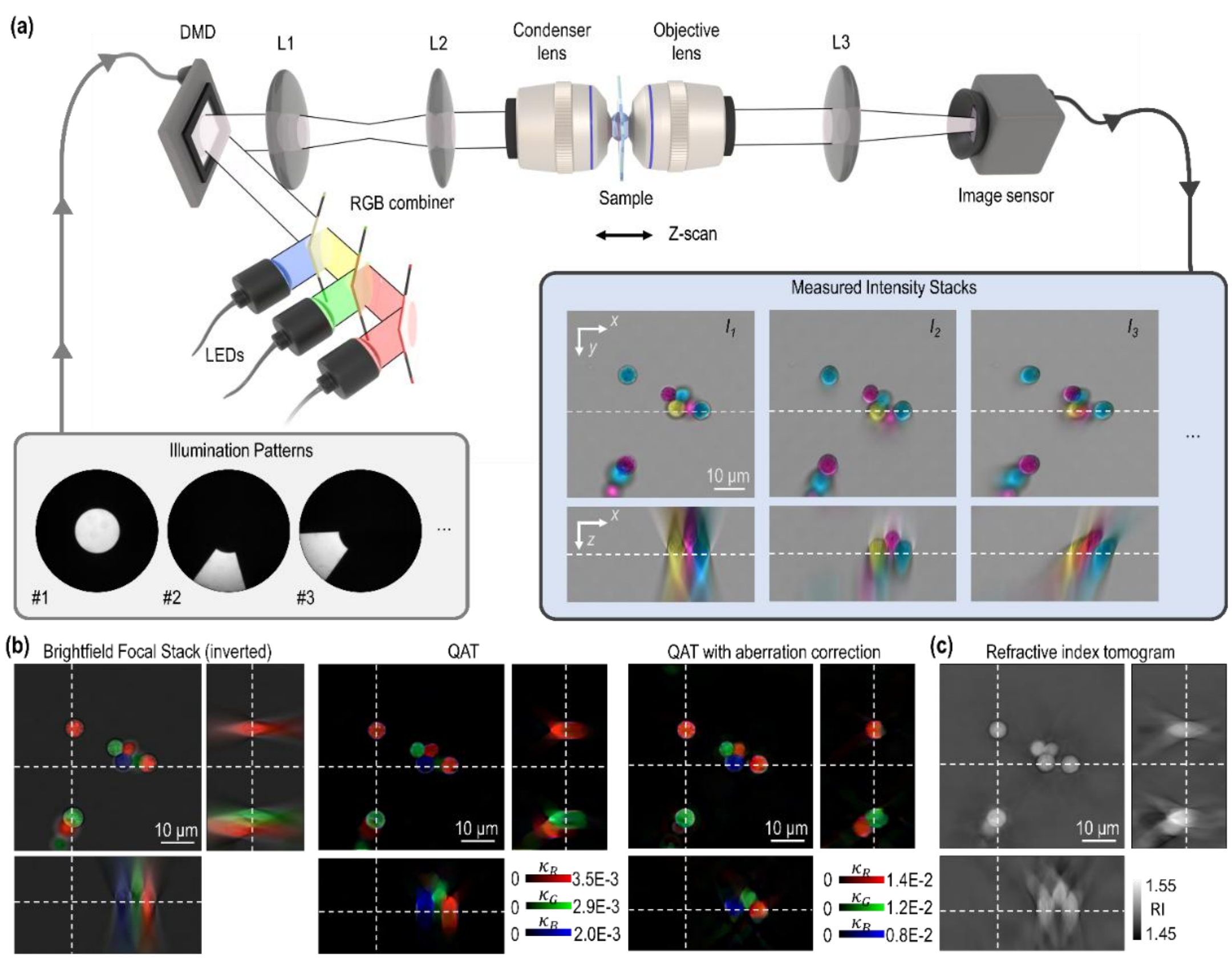

**Figure 1 | Principle and validation of QAT.** (a) Optical implementation. A DMD placed at the condenser pupil sequentially projects programmed illumination patterns under red (624 nm), green (520 nm), and blue (455 nm) LED illumination. Transmitted light is collected by a high-NA objective, and through-focus intensity stacks are recorded by axial scanning of the sample. (inset, left), Representative pupil illumination patterns generated by the DMD (three shown out of 8), which together form an effective partially coherent brightfield illumination for absorption reconstruction. (inset, right), Measured RGB brightfield focal stacks acquired at different axial positions ($I_1$–$I_3$ shown). (b) QAT reconstruction results for a mixture of cyan, magenta, and yellow (CMY) laser printer toner particles, which selectively absorb red, green, and blue light, respectively. (c) Refractive index tomogram reconstructed from multi-illumination measurements using the green LED.

### Label-free imaging of melanin in melanoma cells

QAT resolves the three-dimensional distribution of endogenous melanin in living melanoma cells without exogenous labels. We applied QAT to B16-BL6 murine melanoma cells, which accumulate melanin — a strongly absorbing pigment central to pigmentation, photoprotection, and disease — with pronounced spatial heterogeneity that makes them a natural test case for volumetric absorption imaging.

In conventional brightfield microscopy, melanin-containing regions are visible as dark features; however, the observed contrast represents a mixture of absorption, phase retardation, and out-of-focus contributions which can be hard to differentiate. Representative brightfield focal stacks of melanoma cells (Fig. 2a, left column) illustrate this limitation: melanin appears axially elongated, its depth distribution is ambiguous, and its out-of-focus phase artifacts can be mistaken for absorption. As a result, conventional brightfield imaging does not allow quantitative or 3D interpretation of intracellular melanin organization.

QAT reconstruction of the extinction coefficient transforms this picture (Fig. 2a, middle column). By strongly suppressing phase-induced contrast and out-of-focus background, QAT enhances absorption-dominant contrast and recovers the three-dimensional distribution of melanin within individual melanoma cells. Orthogonal slices reveal sharply confined absorption signals corresponding to melanin-rich subcellular regions, enabling accurate depth localization and volumetric mapping. For comparison, RI tomograms reconstructed from the same datasets are shown in Fig. 2a (right column). RI imaging delineates overall cellular morphology, including cell boundaries and organelle-rich regions, but does not selectively highlight melanin. Fig. 2b further highlights the ability of QAT to capture variability in intracellular pigments among melanoma. Absorption signals range from localized subcellular domains to larger volumetric distributions.

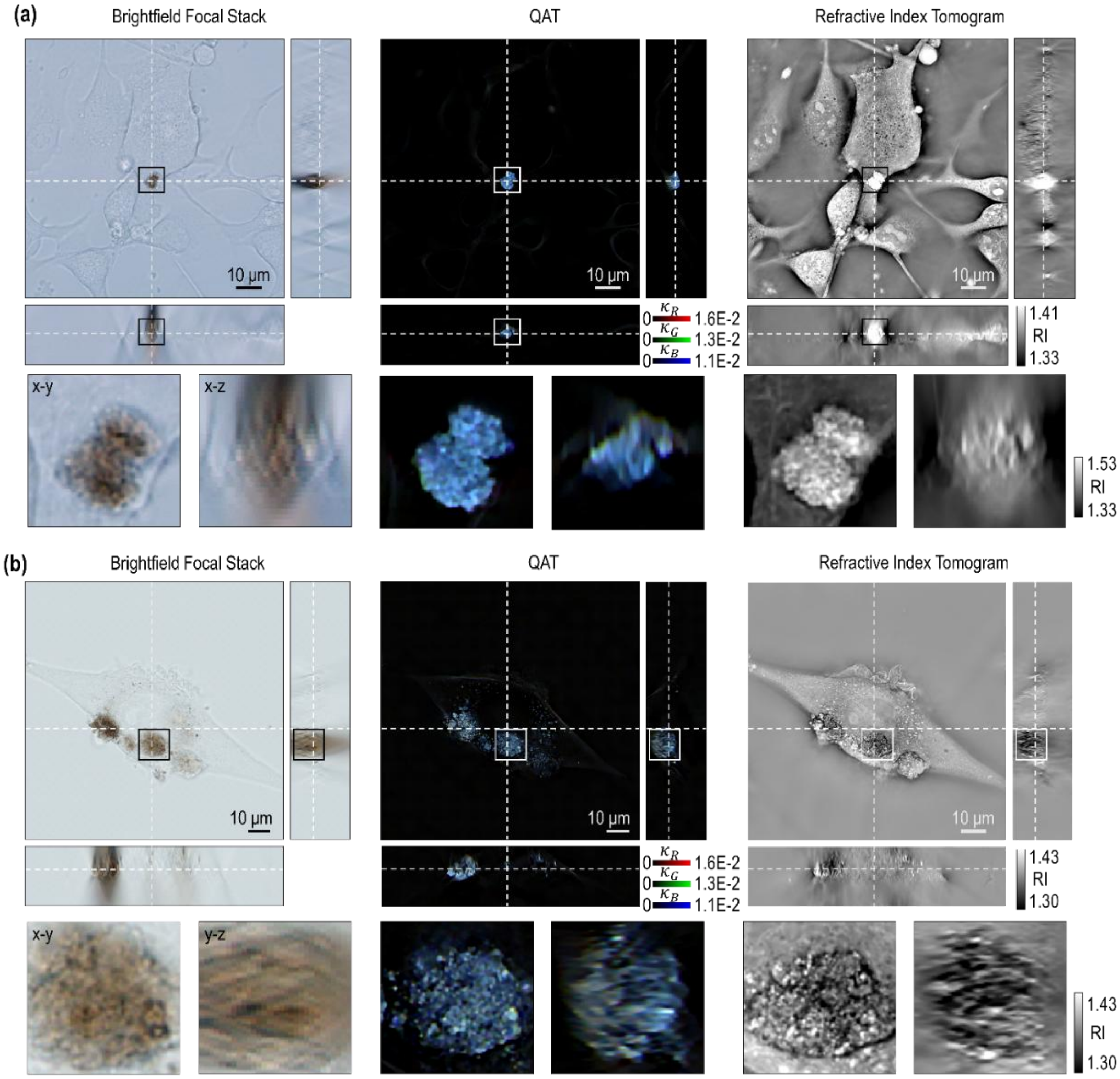

**Figure 2 | Quantitative absorption tomography of melanoma cells.** (a) Representative cells with localized intracellular pigment. Left column: Raw brightfield images. Middle column: QAT reconstructions provide enhanced axial sectioning and accurate 3D localization of absorbing structures. Orthogonal slices (x–y, x–z, and y–z; white crosshairs) highlight the volumetric distribution of pigment with strong suppression of out-of-focus background. Right column: Corresponding RI tomograms of the same cells reveal

overall cellular morphology but lack specificity to pigment accumulation, underscoring the complementary contrast provided by absorption imaging. (b) Additional examples showing more prominent, spatially extended absorption features. Insets display zoomed views of reconstructed absorption volumes, emphasizing subcellular pigment organization.

A major advantage of QAT is its ability to image endogenous absorbers without exogenous labels, enabling long-term live-cell measurements with minimal perturbation or phototoxicity. We leveraged this label-free capability to perform time-lapse, 3D monitoring of melanogenesis. We applied QAT to B16-BL6 murine melanoma cells undergoing forskolin-induced melanogenesis and acquired volumetric datasets over 24-hours (Fig. 3). This label-free configuration avoids photobleaching, phototoxicity, and molecular interference, which commonly limit long-term fluorescence-based studies of pigment biology[31].

Representative brightfield images acquired at successive time points are shown in Fig. 3 (first row). Although melanin-containing regions appear as dark, brightfield images suffer from mixed phase–absorption contrast and strong axial blur, obscuring both the 3D organization of melanin and its temporal evolution. As a result, identifying melanogenic cells and tracking pigment dynamics over time remains challenging using brightfield microscopy alone.

QAT absorption reconstructions provide a distinct view by strongly suppressing phase-related background signals, thereby enhancing contrast associated with melanin accumulation (Fig. 3, second and third rows). Absorption axial slices and maximum-intensity projections show the gradual emergence of localized absorption signals in a subset of cells, corresponding to melanin accumulation. Over time, the absorption signals exhibit dynamic spatial redistribution along dendritic extensions, consistent with melanin-associated contrast. These observations are consistent with known melanosome transport processes and demonstrate that QAT enables direct, 3D visualization of melanin dynamics in living cells.

Simultaneously reconstructed RI tomograms (Fig. 3, fourth row) provide high-contrast visualization of overall cell morphology. However, RI contrast alone does not distinguish melanogenic cells from non-melanogenic neighbors, underscoring the limited functional specificity of phase-based imaging for pigment-related processes.

Overlaying QAT absorption maps on RI tomograms (Fig. 3, fifth row) highlights the complementary nature of the two contrasts. Absorption signals are confined to melanin-rich subcellular regions, while RI maps provide the surrounding cellular context. Magnified views further reveal subcellular-scale pigment aggregation, transport, and reorganization, emphasizing the ability of QAT to resolve dynamic absorption-driven processes in three dimensions.

All live-cell experiments were performed using incoherent LED illumination at low irradiance (1.3 $mW/cm^2$ measured at the sample plane). For visualization, figures show selected time points at 6-h intervals (data acquired every 20 min). The total optical dose delivered over the 24 h acquisition was substantially lower than that typically used in fluorescence time-lapse imaging. Under these conditions, we observed no detectable photobleaching, morphological abnormalities, or growth arrest, indicating minimal phototoxic perturbation.

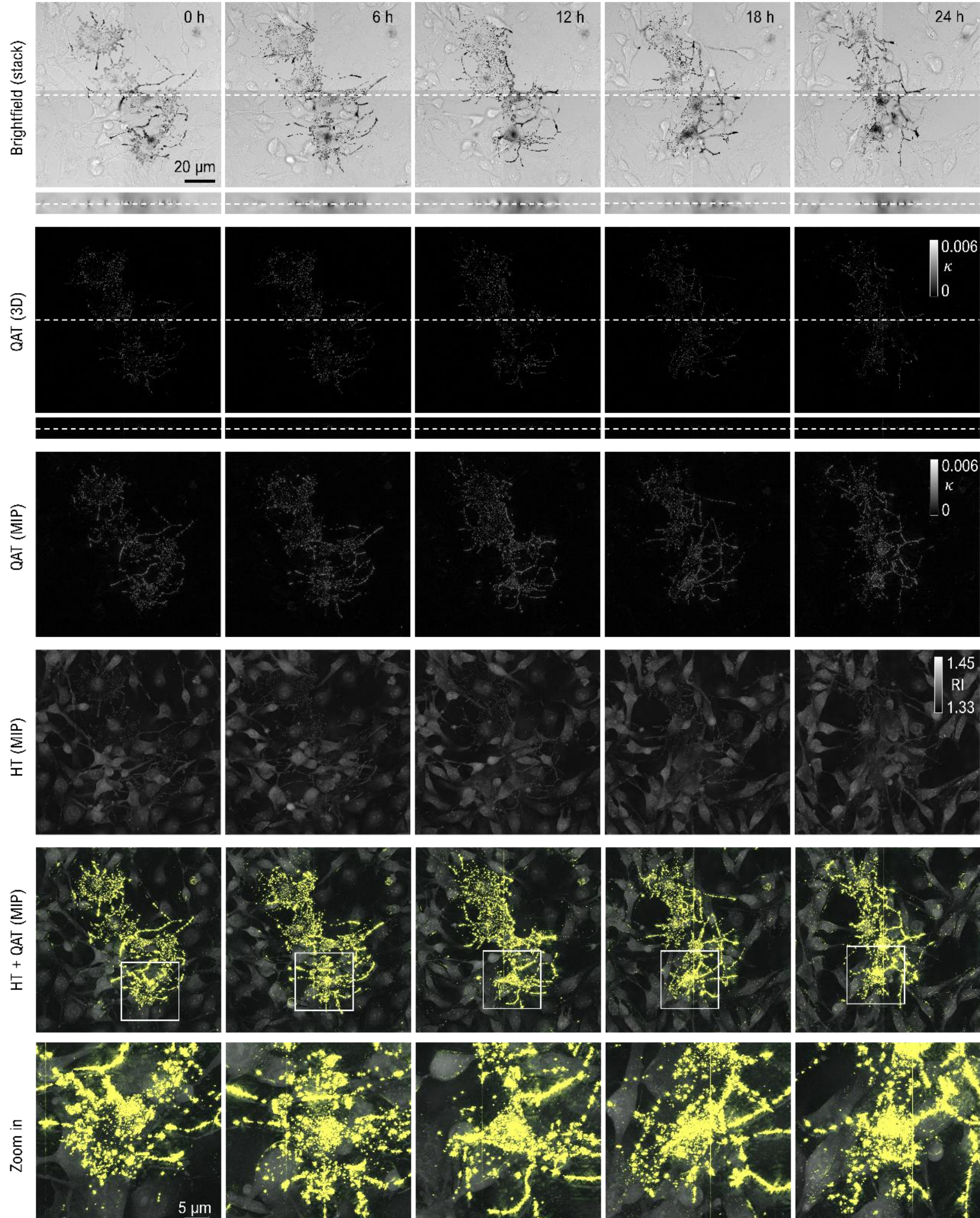

**Figure 3 | Live-cell quantitative absorption tomography reveals dynamic melanogenesis in melanoma cells.** Time-lapse imaging of forskolin-induced melanogenesis in living B16-BL6 melanoma cells over 24 h using quantitative absorption tomography (QAT) (acquired every 20 min; time points displayed at 6-h intervals). Top row: Representative brightfield images at successive time points show weak contrast and substantial axial blur, limiting visualization of pigment dynamics. Second row: Corresponding QAT absorption axial slices highlight melanin-associated absorption with reduced out-of-focus background, enabling clear localization of pigment-related signals across depth. Third row: QAT absorption maximum-intensity projections (MIPs) summarize the volumetric signal and visualize its emergence and redistribution over time with improved contrast. Fourth row: RI tomograms of the same field of view capture overall cell morphology and proliferation but do not distinguish pigment-producing cells. Fifth row: Overlay of QAT absorption (yellow) on RI maps (grayscale) highlights selective melanin accumulation in a subset of cells, primarily along dendritic extensions. Bottom row: Magnified views of the boxed regions illustrate subcellular-scale pigment transport, aggregation, and spatial reorganization over time. QAT enables stable, label-free monitoring of intracellular pigment dynamics over long durations, revealing heterogeneous melanogenic responses that are not discernible from phase-based contrast alone.

### *In vivo* volumetric absorption imaging of living *Petunia hybrida* petal cells

To demonstrate the applicability of QAT beyond animal cells and into plant biology, we performed *in vivo* time-lapse 3D absorption imaging of living *Petunia hybrida* petal cells. Plant petals represent optically complex, pigment-rich tissues containing multiple endogenous chromophores whose spatial organization and dynamics are important for plant function and reproduction, yet remain challenging to visualize in three dimensions under native conditions.

We imaged intact *Petunia hybrida* petals while the flower remained attached to the living plant (Fig. 4a). A single petal was gently positioned between two coverslips on the microscope stage, and hydration was maintained throughout the experiment to preserve physiological viability and RI matching. This minimally invasive configuration enabled repeated acquisition of volumetric focal stacks over time without chemical fixation, labeling, or detachment of the tissue from the organism.

Raw brightfield focal stacks exhibit strong axial blur and limited depth discrimination, obscuring the 3D organization of epidermal cells and their intracellular pigment pools (Fig. 4b). In contrast, QAT reconstruction yields high-resolution, RGB absorption-weighted volumes that resolve individual epidermal cells and their pigment-filled vacuoles throughout the tissue thickness (Fig. 4c). Orthogonal cross-sections reveal sharply localized absorption signals, demonstrating effective axial sectioning in an optically heterogeneous, living plant tissue.

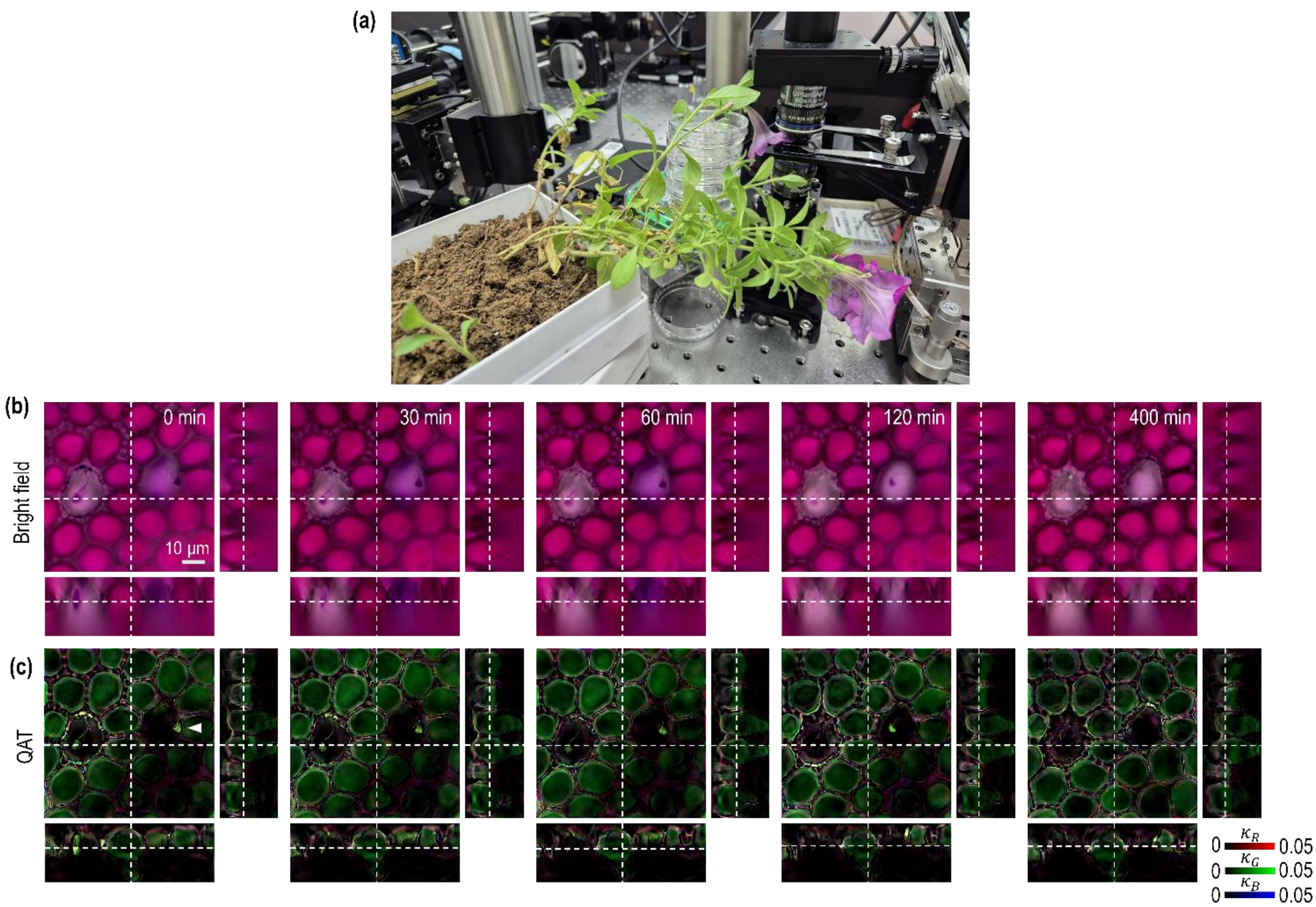

**Figure 4 | *In vivo* quantitative absorption tomography of living *Petunia hybrida* petal cells.** (a) Experimental configuration for in vivo imaging of living *Petunia hybrida* petal cells. The flower remains attached to the plant, while a single petal is gently positioned between coverslips on the microscope stage. (b) Representative raw brightfield focal-stack images of petal epidermal cells. (c) RGB QAT absorption reconstructions of petal epidermal cells shown as orthogonal slices (x–y, x–z, and y–z; white crosshairs). Vacuolar pigment absorption is resolved throughout the tissue thickness with strong axial sectioning. Color-composite absorption volumes enable volumetric quantification of pigment-filled vacuoles and cell-to-cell variability in pigment distribution, going beyond the qualitative contrast available in conventional two-dimensional brightfield images. QAT enables label-free, volumetric absorption imaging of intact, living plant tissue under minimally invasive conditions.

Beyond visualization, the reconstructed absorption volumes provide quantitative, 3D readouts of pigment-associated absorption within petal cells. For example, the volumetric extent, spatial distribution, and integrated absorption of pigment-rich regions can be measured on a cell-by-cell basis from the absorption maps. These volumetric measures facilitate comparison of pigment organization and variability across neighboring cells in a manner that is difficult to achieve from conventional 2D brightfield images alone.

Time-lapse QAT imaging captures dynamic changes in pigment organization within living petal cells. Over successive time points, we observed gradual redistribution of pigment pools in a subset of cells, manifested as changes in the 3D shape and extent of absorption regions.

**Volumetric chromogenic mapping of H&E-stained human tissue**

QAT scales to large-area H&E-stained human tissue sections, reconstructing volumetric chromogenic contrast without serial sectioning. H&E staining has been the cornerstone of histopathology for more than a century, yet pathology assessment has remained largely qualitative and two-dimensional. Hematoxylin selectively stains cell nuclei (purple), while eosin highlights cytoplasmic and extracellular components (pink), together providing a powerful and intuitive visualization of tissue architecture. Despite its central role in clinical diagnosis, H&E-based pathology has remained limited to qualitative, two-dimensional assessment of thin tissue sections, relying heavily on visual interpretation by pathologists. To overcome these limitations, we applied QAT to formalin-fixed, paraffin-embedded (FFPE) human rectal tissue sections stained with H&E, enabling volumetric, quantitative absorption imaging of standard histological specimens.

Unlike conventional brightfield microscopy (Fig. 5a), which integrates out-of-focus contributions along the optical axis, QAT (Fig. 5b) reconstructs the 3D distribution of absorption coefficients, directly corresponding to hematoxylin- and eosin-associated pigments. QAT markedly suppresses axial blur and resolves absorption contrast throughout the tissue thickness. 3D renderings of the reconstructed absorption volumes (Fig. 5c) visualize depth-dependent variations in absorption contrast, which can aid qualitative assessment of microarchitectural organization beyond two-dimensional sections.

To evaluate scalability, we performed tiled QAT imaging and reconstructed a large-scale absorption volume from a large-area rectal tissue section (Fig. 5d). Tile acquisition used the single-illumination absorption-only QAT mode to maximize throughput; refractive-index recovery was not pursued in this dataset. The stitched volume preserves submicron lateral and axial resolution across the entire field of view, maintaining consistent spectral absorption contrast across tiles. Representative subregions illustrate diverse tissue architectures, including fibrous structure (Fig. 5d (i)), nuclear-dense regions (Fig. 5d (ii)), and eosin-rich cytoplasmic domains (Fig. 5d (iii)), all visualized with enhanced axial sectioning and spectral fidelity compared to conventional brightfield images.

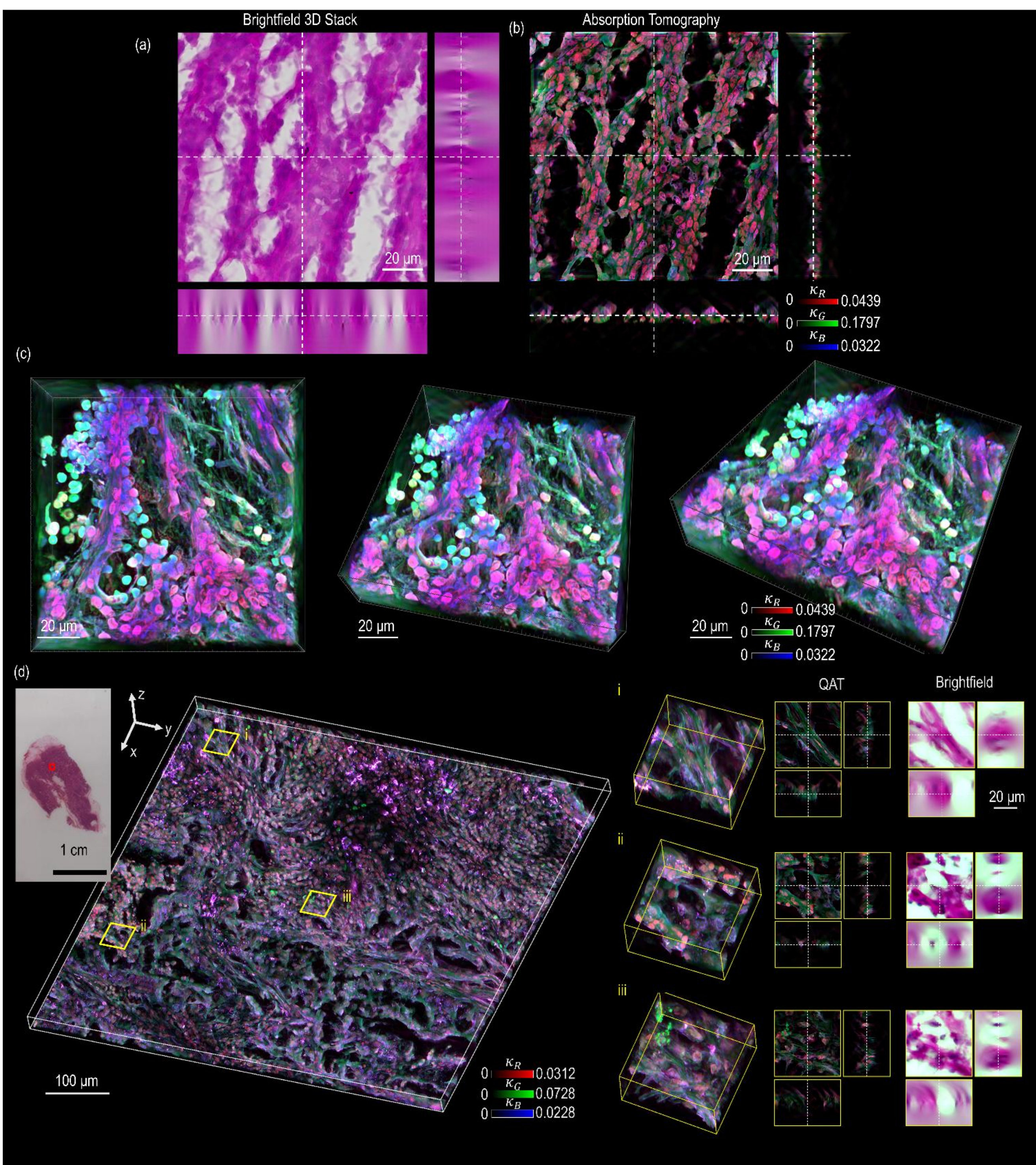

**Figure 5 | Large-scale 3D absorption imaging of H&E-stained human tissue using QAT.** (a) Representative orthogonal slices from conventional brightfield focal stacks and (b) corresponding RGB QAT absorption reconstructions of an H&E-stained human rectal tissue section. (c) 3D maximum-intensity-projection (MIP) renderings of the reconstructed absorption volume, shown from different viewing angles, revealing volumetric organization of tissue microarchitecture. (d) Large-area stitched QAT volume covering a large-scale tissue region (left; inset shows a macroscopic photograph of the section). Three representative subvolumes (i–iii) are highlighted and compared with corresponding conventional brightfield (BF) images (right). QAT preserves spectral absorption contrast and fine structural detail across scales, whereas BF images remain dominated by axial blur.

## Discussions

A key conceptual advance of QAT lies in recognizing that optical absorption can be treated as a linear imaging process under weak-scattering conditions when expressed in logarithmic intensity space. This formulation allows the absorption OTF to be analytically computed and inverted, yielding quantitative 3D extinction coefficient distributions. QAT does not rely on interferometry, coherent illumination, or sample rotation, making it directly compatible with widely used brightfield microscope architectures.

QAT scales to standard H&E-stained human tissue specimens commonly used in histopathological workflows, enabling volumetric reconstruction of chromogenic absorption contrast beyond conventional two-dimensional imaging (Fig. 5). For more than a century, hematoxylin and eosin have served as highly effective pigments for selectively staining nuclei and cytoplasm, respectively; yet pathology assessment has remained largely qualitative and two-dimensional. By reconstructing volumetric, quantitative absorption maps from standard H&E-stained sections, QAT enables 3D visualization of tissue microarchitecture with high-resolution subcellular detail that was previously inaccessible. Beyond improved visualization, the quantitative readout of QAT extends the analytical reach of histopathology. Converting H&E contrast into 3D, quantitative absorption measurements provide objective descriptors of nuclear and cytoplasmic organization that can be systematically analyzed. While clinical validation is beyond the scope of this study, such quantitative descriptors may serve as a foundation for future efforts toward more reproducible and quantitative histopathological assessment, which can potentially reduce inter-observer variability inherent in conventional qualitative pathology.

Our results show that QAT provides effective axial sectioning and spectral specificity that are not accessible with conventional brightfield imaging. Validation experiments using spectrally selective CMY toner particles (Fig. 1) confirmed accurate 3D localization and faithful recovery of wavelength-dependent absorption contrast, establishing QAT as a quantitative imaging modality rather than a contrast-enhancement technique.

At the cellular level, QAT reveals absorption-specific information that is complementary to refractive index tomography. Experiments in B16-BL6 melanoma cells —volumetric mapping of pigment-associated absorption (Fig. 2) and time-lapse imaging of melanogenesis (Fig. 3) — highlight the complementarity of refractive index-based and absorption-based imaging. While refractive index contrast effectively captures overall cellular morphology, it lacks chemical specificity for chromogenic or pigmented structures. In contrast, absorption contrast directly reports on pigment content and spatial distribution. By isolating absorption from mixed brightfield signals, QAT enables unambiguous identification and longitudinal tracking of pigment-producing cells in a label-free manner, demonstrating its utility for functional live-cell studies. Quantitative 3D mapping of melanin without exogenous labels opens new opportunities to study melanogenesis, melanosome transport, intracellular pigment organization, and cell-to-cell variability. More broadly, this work positions absorption tomography as a distinct and previously underexplored modality in biophotonics, complementing phase-based approaches and providing direct access to pathophysiologically important pigments.

The applicability of QAT to intact, optically complex tissues is further demonstrated by in vivo imaging of living *Petunia hybrida* petal cells (Fig. 4). Plant tissues pose substantial challenges due to their thickness, endogenous pigmentation, and refractive index heterogeneity. Nevertheless, QAT resolves 3D pigment distributions and captures time-resolved changes under native conditions while maintaining tissue viability. This capability extends absorption tomography from cell biology to plant biology, opening new avenues for studying pigment biosynthesis, transport, and spatial regulation in living plants.

Despite these advances, several limitations remain. Although QAT strongly suppresses phase-induced intensity variations, it does not claim strict or universal decoupling of absorption and refraction under all imaging conditions. QAT also relies on the weak-scattering approximation and assumes absorption-dominant contrast; highly scattering or strongly refractive specimens may violate these assumptions and introduce reconstruction artifacts. Residual phase–absorption coupling may persist in specimens with strong RI heterogeneity or multiple scattering, in single-illumination configurations, or whenever the accumulated phase delay approaches one radian (Supplementary Note 1); under such conditions, multi-illumination acquisition is required to recover reliable absorption maps. Multi-illumination schemes can decouple phase and absorption contributions but increase acquisition time and computational complexity. Axial resolution is also constrained by the numerical aperture of the optical system and by signal-to-noise limitations inherent to absorption imaging.

Future work may address these challenges through optimized illumination strategies, advanced regularization informed by tissue priors, and hybrid reconstructions that jointly exploit absorption and refractive index information. Integration with machine learning could further enhance robustness and enable automated analysis[32], including segmentation of subcellular organelles[33], classification of cell states[34], and virtual staining[35,36]. Extending QAT to chromogenic immunohistochemistry, multiplexed stains, and whole-slide imaging represents a promising direction toward 3D spatial biology and quantitative pathology[3,37,38]. Parallel advances in non-interferometric brightfield tomography and partially coherent intensity inverse problems[15,18,19] will likely shape the algorithmic backbone of these extensions.

## Conclusion

In this work, we introduced QAT, a computational imaging framework that enables 3D, quantitative reconstruction of absorption contrast from intensity image stacks. By modeling absorption image formation in logarithmic intensity space and applying 3D deconvolution, QAT transforms stain-based brightfield microscopy from qualitative two-dimensional imaging into quantitative volumetric absorption imaging with subcellular resolution.

QAT is broadly deployable: it operates on standard slide-mounted specimens, requires no interferometry or sample rotation, and is fully compatible with conventional brightfield microscope hardware combined with axial scanning. This accessibility allows quantitative 3D absorption imaging of both histochemical tissues and intrinsically pigmented samples without altering established workflows. By enabling volumetric, spectrally resolved absorption mapping in stained human tissue, living cells, and intact plant tissue, QAT establishes absorption tomography as a complementary imaging axis alongside phase- and fluorescence-based methods. Converting chromogenic contrast into 3D quantitative measurements opens a path toward more objective and reproducible analysis of tissue architecture in histopathology.

Overall, QAT provides a physically grounded and widely deployable route to 3D absorption imaging, bridging conventional brightfield microscopy and quantitative volumetric analysis, and laying a foundation for future advances in spatial biology and quantitative pathology.

## Methods

### Imaging setup

QAT can be implemented on conventional brightfield microscopes equipped with axial scanning. In this study, we employed a custom-built brightfield transmission microscope to acquire through-focus intensity stacks under Köhler illumination. Axial scanning of the sample was used to encode 3D absorption information.

To enable programmable angular illumination and, when required, simultaneous reconstruction of absorption and refractive index, a DMD (DLP® LightCrafter™ 4500, Texas Instruments) was placed at the condenser pupil plane. The DMD generated predefined illumination patterns that controlled the pupil filling of the condenser, allowing flexible realization of both single-illumination and multi-illumination acquisition schemes.

LEDs at central wavelengths of 624 nm (red), 520 nm (green), and 455 nm (blue) (M625L4, M530L4, M455L4; Thorlabs) were used as illumination sources. The illumination was delivered to the sample through a high–numerical-aperture water-immersion condenser (UPLSAPO60XW, NA 1.2; Olympus). Transmitted light was collected by a matched 60× water-immersion objective lens (UPLSAPO60XW, NA 1.2; Olympus) and relayed onto a monochrome CMOS camera (ORX-10G-71S7M-C, Oryx; pixel size 4.5 μm) via an achromatic tube lens.

Axial scanning was performed using a piezoelectric translation stage (Q-545, Physik Instrumente). For large-area volumetric imaging, a motorized XY translation stage (LNR25M, Thorlabs) was used, with adjacent tiles acquired using a 25 μm lateral overlap to facilitate stitching.

For live-cell melanoma experiments, data were acquired using a HT platform (HT-X1, Tomocube Inc., Republic of Korea) optimized for long-term live-cell imaging. This system employs a 450 nm LED illumination source and a DMD-based structured illumination scheme, enabling acquisition protocols compatible with QAT reconstruction.

### Sample Preparation

**Validation phantoms.** Spectrally selective absorption phantoms were prepared using cyan, magenta, and yellow (CMY) laser printer toner particles (HP 218A). The particles were dispersed on glass slides and sealed with a coverslip using mounting medium. These phantoms were used to validate wavelength-specific absorption reconstruction, three-dimensional localization accuracy, and RGB composite visualization.

**Melanoma cells.** B16-BL6 murine melanoma cells were cultured in DMEM supplemented with 10% FBS and 1% penicillin-streptomycin at 37 °C and 5% $CO_2$. No serum starvation was performed. Melanogenesis was induced by treatment with 20 μM forskolin (DMSO vehicle control), which was maintained throughout imaging.

Cells were seeded onto glass-bottom dishes and allowed to adhere overnight. Live-cell imaging was performed using an HT-X1 platform (Tomocube Inc.) with integrated temperature (37 °C) and $CO_2$ control. Volumetric datasets were acquired from the same field of view every 20 min for 24 h at an illumination intensity of 1.3 mW/cm$^2$. No detectable morphological changes or detachment were observed during imaging.

**Petunia flower petal.** *Petunia hybrida* petals were imaged in *vivo* while the flower remained attached to the living plant. A single petal was gently positioned between two glass coverslips and kept hydrated throughout imaging to maintain physiological viability. The plant was returned to normal growth conditions after imaging, with no visible damage observed.

**Human rectal tissue.** Formalin-fixed, paraffin-embedded (FFPE) human rectal tissue samples were sectioned to thicknesses of 10–20 µm to preserve three-dimensional tissue architecture. Sections were stained with hematoxylin and eosin (H&E) using standard histological protocols, deparaffinized, and mounted in a refractive index–matched mounting medium. Prepared sections were placed between a glass slide and coverslip for brightfield imaging.

**Data and Code Availability**

The data that support the findings of this study are available from the corresponding author upon reasonable request. Custom code used for analysis and reconstruction is available from the corresponding author upon reasonable request.

**Acknowledgements**

This work was supported by National Research Foundation of Korea grant funded by the Korea government (MSIT) (RS-2024-00442348, RS-2022-NR068141), Korea Institute for Advancement of Technology (KIAT) through the International Cooperative R&D program (P0028463), the Korean Fund for Regenerative Medicine (KFRM) grant funded by the Korea government (the Ministry of Science and ICT and the Ministry of Health & Welfare) (RS-2024-00332454), Commercialization Promotion Agency for R&D Outcomes (COMPA) funded by the Ministry of Science and ICT (MSIT) (RS-2024-00440577), and the Samsung Research Funding Center of Samsung Electronics under Grant (SRFC-IT1401-08).

**Competing Interests**

Y.K.P. has financial interests in Tomocube Inc., a company that commercializes HT products. All other authors declare no competing interests.